\documentclass[12pt]{article}
\usepackage{authblk}
\usepackage[numbers,sort&compress]{natbib}
\usepackage[margin=1in]{geometry}
\usepackage{chemformula} % Formula subscripts using \ch{}
\usepackage[T1]{fontenc} % Use modern font encodings
\usepackage{graphicx}
\RequirePackage{mathpazo}         % Font of the document
\DeclareMathSizes{10}{10}{8}{7.5}
\RequirePackage[T1]{fontenc}      % Encoding of the font
\RequirePackage{fix-cm} 
\usepackage{float}

\usepackage{setspace} 
\begin{document}
\title%[An \textsf{achemso} demo]
{\textbf{Inter-flake transport and humidity response of \ch{Ti_3C_2T_x} MXene at the nanoscale}}
\date{}
\author[1]{Oriane de Leuze}
%\affiliation[]{IMCN, UCLouvain, Louvain-La-Neuve, Belgium}
%\email{oriane.deleuze@uclouvain.be}
\author[2]{Maxime Berthe}
%\affiliation[]{IEMN, CNRS, Université de Lille, Villeneuve d’Ascq, France}
\author[1]{Sophie Hermans}
%\affiliation[]{IMCN, UCLouvain, Louvain-La-Neuve, Belgium}
\author[1]{Benoît Hackens}
%\affiliation[]{IMCN, UCLouvain, Louvain-La-Neuve, Belgium}
%\email{benoit.hackens@uclouvain.be}
\affil[1]{IMCN, UCLouvain, Louvain-La-Neuve, Belgium}
\affil[2]{IEMN, CNRS, Université de Lille, Villeneuve d’Ascq, France}

  \maketitle
%Nanoscale Transport and Humidity Response across Ti₃C₂Tₓ Flake Junctions

%% MESSAGE DE L'ARTICLE %%
%Idée titre: Nanoscale study of inter-flake charge transport in Ti3C2Tx MXene
%We show 
%- potential drop localized @junction
%- Conductive MXene path == isopotentials separated by junction

\begin{abstract}

\noindent Understanding charge transport in networks of two-dimensional crystals is essential for developing reliable applications such as chemiresistors or electromagnetic shields. For this purpose, intra- and inter-flake contributions to the network resistance must be disentangled. MXenes, such as \ch{Ti_3C_2T_x}, are prime examples of 2D crystals often employed as thin networks of interconnected flakes deposited on substrates to realize functional devices. While a significant number of studies focused on transport in individual MXene flakes, inter-flake transport remains scarcely explored. Here, we demonstrate that charge transport in multi-flake conductive paths of \ch{Ti_3C_2T_x} is dominated by interflake junctions and provide quantitative estimates of junction resistances. Scanning probe measurements reveal that in a MXene multi-flake conductive path, individual flakes behave as isopotential domains, since the voltage drop is localized precisely at the inter-flake junctions. We further investigate the chemiresistive response to humidity at the single flake, multi-flake and flake network scale, evidencing the leading impact of junctions on sensing kinetics. These findings underline the crucial role of junctions in charge transport and sensing capabilities of MXenes.
%leading

%Isopotentials
%Dominance of Rjunct
\end{abstract}

\section*{Introduction}

Scalability has been an ongoing challenge since the birth of 2D crystals with exfoliated graphene. Now, solution-processed layers of 2D crystals are increasingly used in electronics as they comply with standard microfabrication techniques and are compatible with various substrates, from silicon to textile. Such 2D layers can be used in flexible electronics, sensors, and logic devices among others \cite{tang2024solution}. The main deposition techniques that comply with solution-processed 2D crystals include inkjet printing, drop casting, spray-coating and spin coating, yielding thin films formed by networks of overlapping flakes \cite{gabbett2024quantitative}\cite{etchingMXene}. %peut être détailler + ici et mieux valoriser la citation gabbett 
These layers are shown to be very disordered as flake stacking and flake orientation are uncontrolled. Also, flake size can be rather disperse in a suspension, increasing the random, fragmented nature of the deposited layer. It is therefore challenging to understand charge transport in those types of layers and to tune their properties depending on the target application of the device. A way to gain more insight about charge transport in the case of nanosheet networks is by considering the network as many well-defined conductive paths in parallel, an approach used by Gabbett et al. to derive a model describing network resistivity and mobility as functions of network properties such as flake thickness and channel length \cite{Gabbett}. In this approach, each conductive path consists of a series of resistors including nanosheet resistance, $R_{NS}$ and junction resistance between the nanosheets, $R_{J}$. This flake-junction distinction is a key-element in how to approach charge transport in 2D materials, as it underlines the need to understand not only intra-flake but also inter-flake transport mechanisms \cite{coleman2025decoupling}. Extracting meaningful data on junctions is not trivial, due to the many unknowns, such as the voltage drop profile in the vicinity of the junction, the configuration of current paths, the particular stacking configuration, the role of functionalization groups and defects at the edge of the flakes, as well as the nanoscale dimensions. Typically, such scales can be reached with scanning probe microscopy tools, but the problem complexity calls for an approach involving different scanning probe modes.\\

\noindent
In this work, we focus on the case of \ch{Ti_3C_2T_x}, the most widely studied and well-known MXene to date. MXenes are a family of 2D transition metal carbides, nitrides and carbonitrides. They are obtained by selective etching and delamination of a 3D lamellar MAX precursor \cite{NaguibMXenes}. There is a considerable interest in MXenes due to their tunable chemical and electronic properties. Indeed, their surface chemistry is very rich and can be tuned depending on the synthesis technique for example \cite{schied2024reactivity}. The surface groups are noted \ch{T_x} in the general MXene formula, which is \ch{M_{n+1}X_nT_x}, where 'M' represents the transition metal, 'X' is either carbon or nitrogen and 'n' = 1, 2 or 3. Also, the wide range of possible combinations of different transition metals and functional groups makes MXenes a highly versatile material family. \ch{Ti_3C_2T_x} was the first MXene to be synthesized in 2011 and shows now great promise in the fields of flexible electronics, electromagnetic interference shielding, energy storage, and sensing \cite{CarbonEMIShield}\cite{NatureEnergyStorage}\cite{NatureEMIShield}\cite{li2023toward}\cite{de2025anisotropic}. In those applications, MXenes are mostly used as 'films'. Barsoum and Gogotsi insist on the importance of decoupling inter- and intra-flake charge transport in order to gain more control on the electronic properties of those films \cite{roadblocks}. Also, a distinction has to be made between spin-coated films and those which are drop-cast or filtered, as the alignment between the flakes in spin-coated films shows less variability in the quantitative results from one work to another, regarding electrical properties such as conductivity and mobility. In view of this, it is even more interesting to consider the problem at flake-scale. Yet, such scales are challenging to reach given the small size of the flakes, requiring scanning probe microscopies or microfabrication processes to pattern contacts on the areas of interest \cite{nirmalraj2009electrical}\cite{lipatov2023metallic}\cite{sahare2022assessment}. Several studies have been carried out on individual flakes, both at room and low temperature, shedding light on transport mechanisms in \ch{Ti_3C_2T_x} \cite{Hemmat}\cite{Shekhirev_ultralarge}\cite{LIPATOV20211413}\cite{lipatov2023metallic_REV}\cite{lipatov2023metallic_REV}\cite{Sang}. There is an agreement in literature about the all-time metallic nature of \ch{Ti_3C_2T_x} and about the fact that the individual flake conductivity will directly be affected by the amount of defects and oxidation state of the flake \cite{lipatov2023metallic}\cite{lipatov2016effect_REV}\cite{LIPATOV20211413}. Although the understanding of transport in individual flake is increasing, both in the monolayer and multilayer case, studies that include inter-flake transport are very scarce. A recent work evaluates the layer-dependent sensing characteristics of MXenes at the flake-size and includes both individual flakes and junctions between flakes, shedding light on specific and contrasting sensing mechanisms in both cases \cite{loes2024layer}. Such a pioneering result further underscores the need for flake-scale experimental characterization as a tool to develop and understand MXenes-based electronic applications. \\ %On the other hand, nanoscale characterization of inter-flake transport is rather challenging to perform so a limited amount of works have been published in this direction.\\ %XXX

\noindent
In this context, we leveraged different methods to study charge transport at the nanoscale both within individual few-layer \ch{Ti_3C_2T_x} flakes and across the junction between flakes. Both electrical-mode scanning probe microscopies and measurements on patterned devices are combined in this work to reach quantitative conclusions and local insights. Lastly, electrical responses to humidity at the intra-flake, inter-flake, and flake network scales reveal different characteristic time scales, relevant in the context of sensing applications involving MXenes.
%in the context of electronic applications involving MXenes, electrical response to humidity variation is shown for the individual flake, for junctions between flakes and for a network of flakes.
%Interest in network characterization
%Loes
%Importance of 4-contact meas to be quantitative
%Intercalation
%KPFM
%Transport -> citer barsoum, hurand etc 

%2D charge transport -> citer piatti2021charge -> 2D 
%We find that charge transport in printed few-layer MXene and MoS2 devices is dominated by the intrinsic transport mechanism of the constituent flakes: MXene exhibits a weakly localized 2D metallic behaviour at any temperature

\section*{Results and discussion}
\begin{figure}[H]
    \centering
    \includegraphics[width=\linewidth]{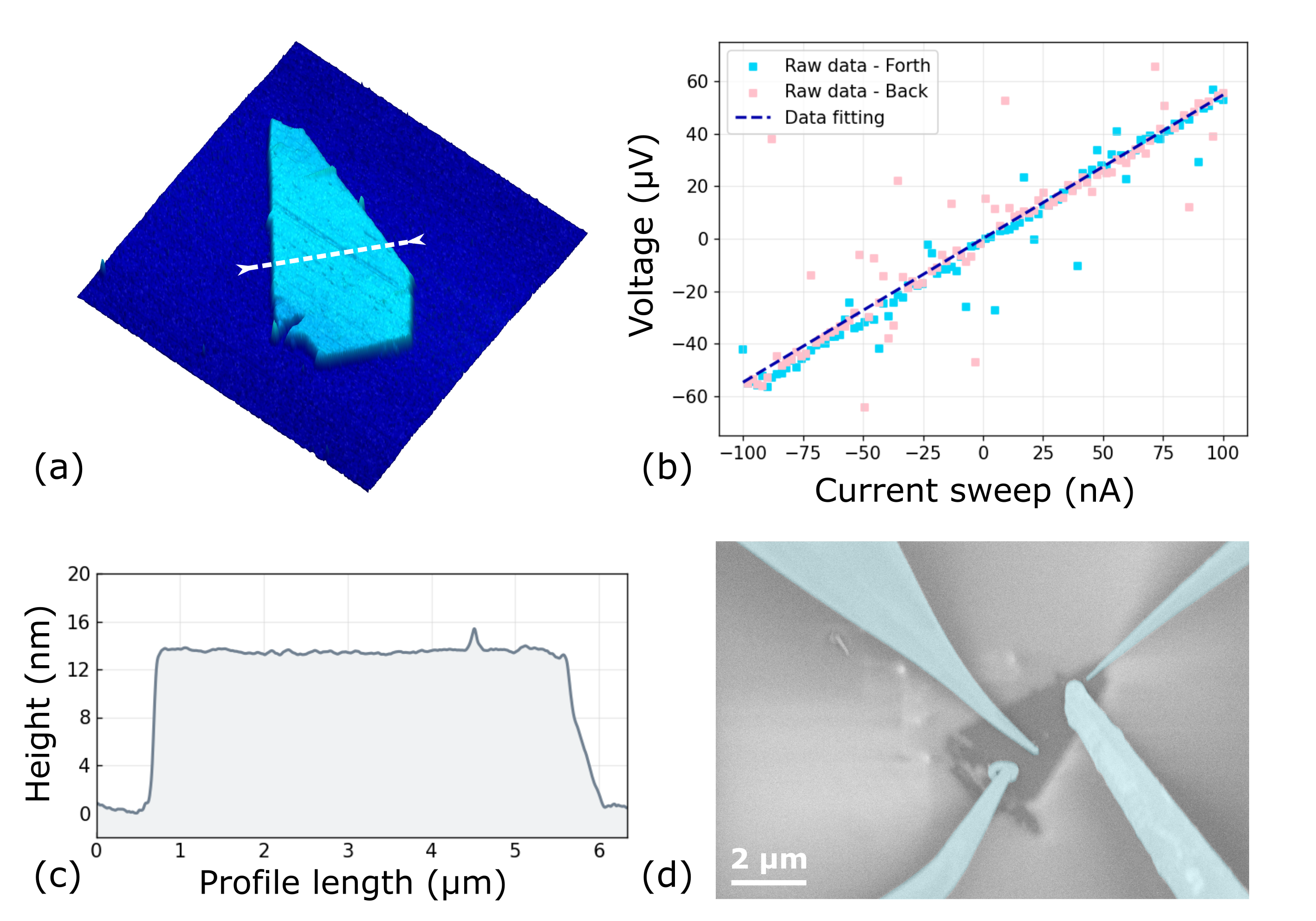}
    \caption{(a) 3D view of the AFM topography of a \ch{Ti_3C_2T_x} flake deposited on a Si/\ch{SiO_2} substrate, obtained in tapping mode. (b) I-V characteristic obtained by 4-probe measurement on the flake shown in (a). (c) Height profile along the white dotted line in (a). (d) SEM image of the \ch{Ti_3C_2T_x} with the tungsten probes placed in the 4-contact measurement configuration.}
    \label{FIG1}
\end{figure}

\paragraph{4-probe STM - Individual flake}
A multiple-probe Scanning Tunneling Microscope (STM) is used to perform 4-contact measurements on individual and multiple \ch{Ti_3C_2T_x} few-layer flakes, lying on an insulating substrate (\ch{SiO_2}-covered Si) as illustrated in Figure \ref{FIG1}. High reliability of the data and stability between different measurements are obtained thanks to the ultrahigh vacuum (UHV) environment of the STM, as well as the accuracy of tip-flake contact position at the nanometer scale and the inter-tip distance that can reach values below 100 nm. This is allowed by the sharp geometry of the tips that have a nm-sized apex and the controlled displacement of the probes carried out under scanning electron microscopy (SEM) visualization. %Given the sharp geometry of the tips that have an apex radius below 100 nm and their controlled displacement under SEM visualization, the accuracy of tip-flake contact position can reach values below 100 nm. This considered along with the UHV atmosphere in the STM chamber ensures high reliability of the data as well as high stability between two different measurements. 
The first experiment carried out is a 4-contact measurement on an individual flake with a homogeneous topography, as depicted on Figure \ref{FIG1} (a), that shows a 3-D view of the AFM topography of the flake. Figure \ref{FIG1} (c) displays the thickness profile extracted along the white dotted line in (a), confirming the topography homogeneity. The four STM probes are positioned on the flake along a straight line as shown on the SEM image in Figure \ref{FIG1} (d). In this configuration, the current is applied between the two outer probes and the voltage drop measured between the two inner probes, whose spacing is 1.8 µm. The flake-resistivity can then be extracted from the current-voltage characteristics that are shown in Figure \ref{FIG1} (b). Those are linear and show no hysteresis, as expected for \ch{Ti_3C_2T_x} given its metallic nature. Knowing the spacing between the probes $d$, the in-plane resistivity value of the flake is obtained by matching the experimental results with a COMSOL finite element simulation: $\rho_{xx}$ = 6.06 $\Omega \cdot \mu m$. An image of the simulated potential distribution on the flake is shown in the supplementary information S1. The use of a finite element simulation is needed here as the size of the flake is in the same range as the probe spacing, which prevents the use of conventional 4-point probe expressions to compute the flake resistivity. The value obtained with the current approach is in agreement with a previous work that reports resistivity values for individual few-layer \ch{Ti_3C_2T_x} flakes in ambient conditions \cite{de2025anisotropic}. Those values are represented on a graph along with flake thickness in the supplementary information S1. This first result validates the 4-probe STM approach to extract quantitative electrical measurements on \ch{Ti_3C_2T_x} flakes and suggests that the in-plane resistivity of \ch{Ti_3C_2T_x} does not vary significantly in an UHV environment, compared to ambient air conditions. Contact resistance is also measured by comparing 2- and 4- contact measurements and is about 2.5 k$\Omega$ in this case, which demonstrates the necessity to perform 4-contact measurements. More details are shown in the supplementary materials S2. \\

\paragraph{4-probe STM - Conductive path} 

\begin{figure}[H]
    \centering
    \includegraphics[width=\linewidth]{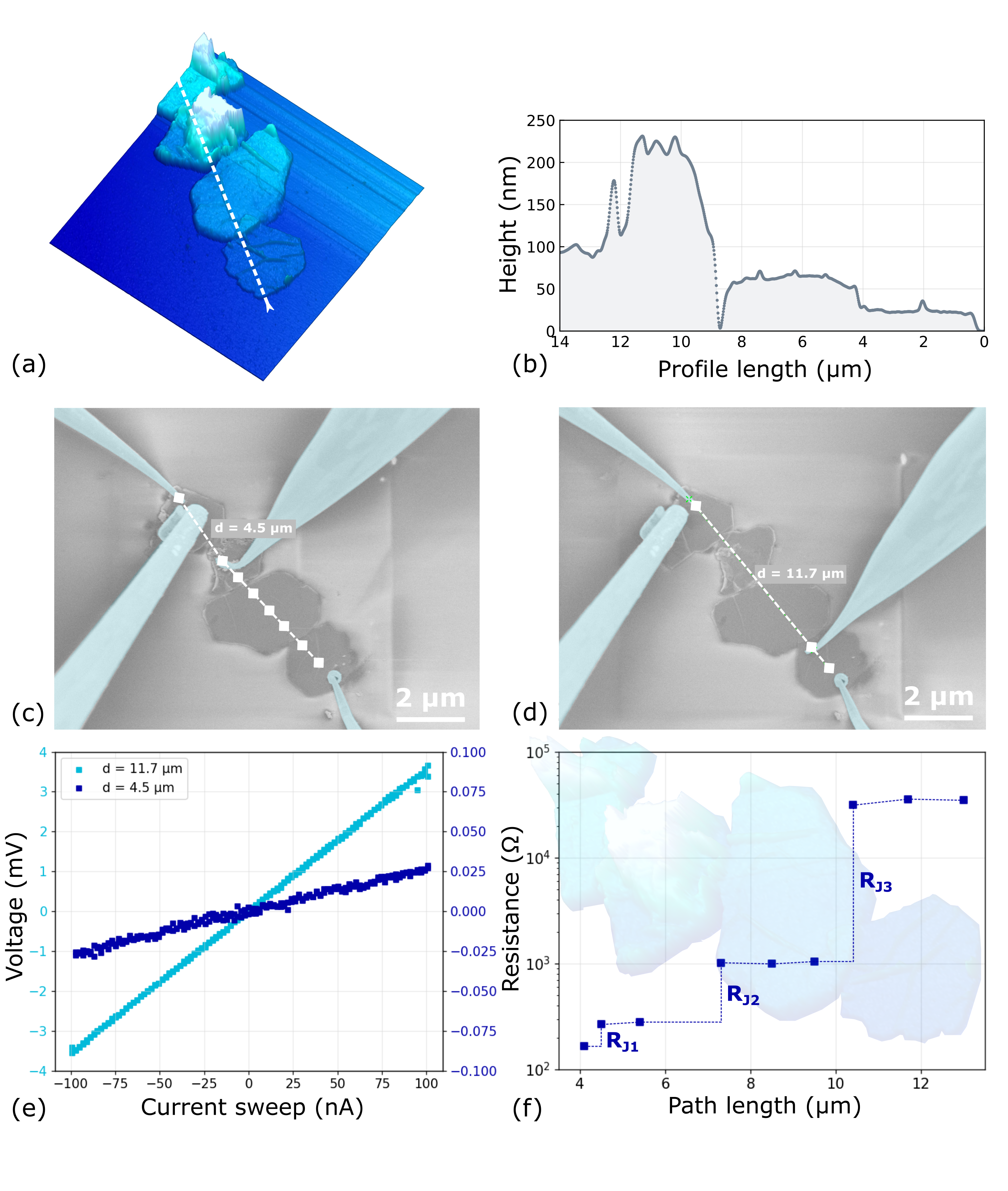}
    \caption{(a) 3D view of the AFM topography of a conductive path formed by \ch{Ti_3C_2T_x} flakes deposited on a Si/\ch{SiO_2} substrate, obtained in tapping mode. (b) Height profile along the white dotted line in (a). (c-d) SEM images of the 4-probe measurement along the conductive path, for two different measurement points, with the distance between the measuring probes being 4.5 and 11.7 µm, respectively. (e) I-V characteristics for both measurements points shown in (c) and (d). Note that the vertical (voltage) scale for the light blue curve (left axis) is orders of magnitude larger than for the dark blue curve (right axis). (f) Electrical resistance along the conducting path as a function of the distance between the measuring probes.}
    \label{FIG2}
\end{figure}

Following intra-flake measurement, we consider the case of a conductive path made of four overlapping individual MXene flakes. This configuration is illustrated in the 3-D view of the topography of the sample on Figure \ref{FIG2} (a) and its thickness profile in Figure \ref{FIG2} (b) along the white dashed line in Figure \ref{FIG2} (a). The conductive path established through these four flakes allows to study each junction separately with the four-probe STM. Hence, four-contact measurements were carried on along the conductive path at different locations, as showcased in Figure \ref{FIG2} (c) and (d), where the indicated distance corresponds to the spacing between measuring probes. The two outer probes are applying a fixed current through the conductive path and only one of the measuring probes is displaced for each measurement point. Similarly to the measurement on an individual flake, an IV curve is acquired for every measurement point. Figure \ref{FIG2} (e) displays the IV characteristics corresponding to the two measurements configuration shown in  the SEM micrographs on Figure \ref{FIG2} (c) and (d). Both are linear, which means that electronic transport through flake junctions is still ohmic and an increase of the resistance is observed with the increase of the distance between the measuring probes. Comparing data obtained when varying the length of the section of the conductive path considered, it is possible to assess this increase of resistance in more detail. Indeed, as presented on Figure \ref{FIG2} (f), the resistance increase occurs by steps corresponding to the junctions between the flakes. Hence, the junction resistances $R_{J1}$, $R_{J2}$ and $R_{J3}$ are the main contributions to the total resistance measured along the conductive path. \\

\paragraph{4-probe STM - Potentiometry}

\begin{figure}[H]
    \centering
    \includegraphics[width=\linewidth]{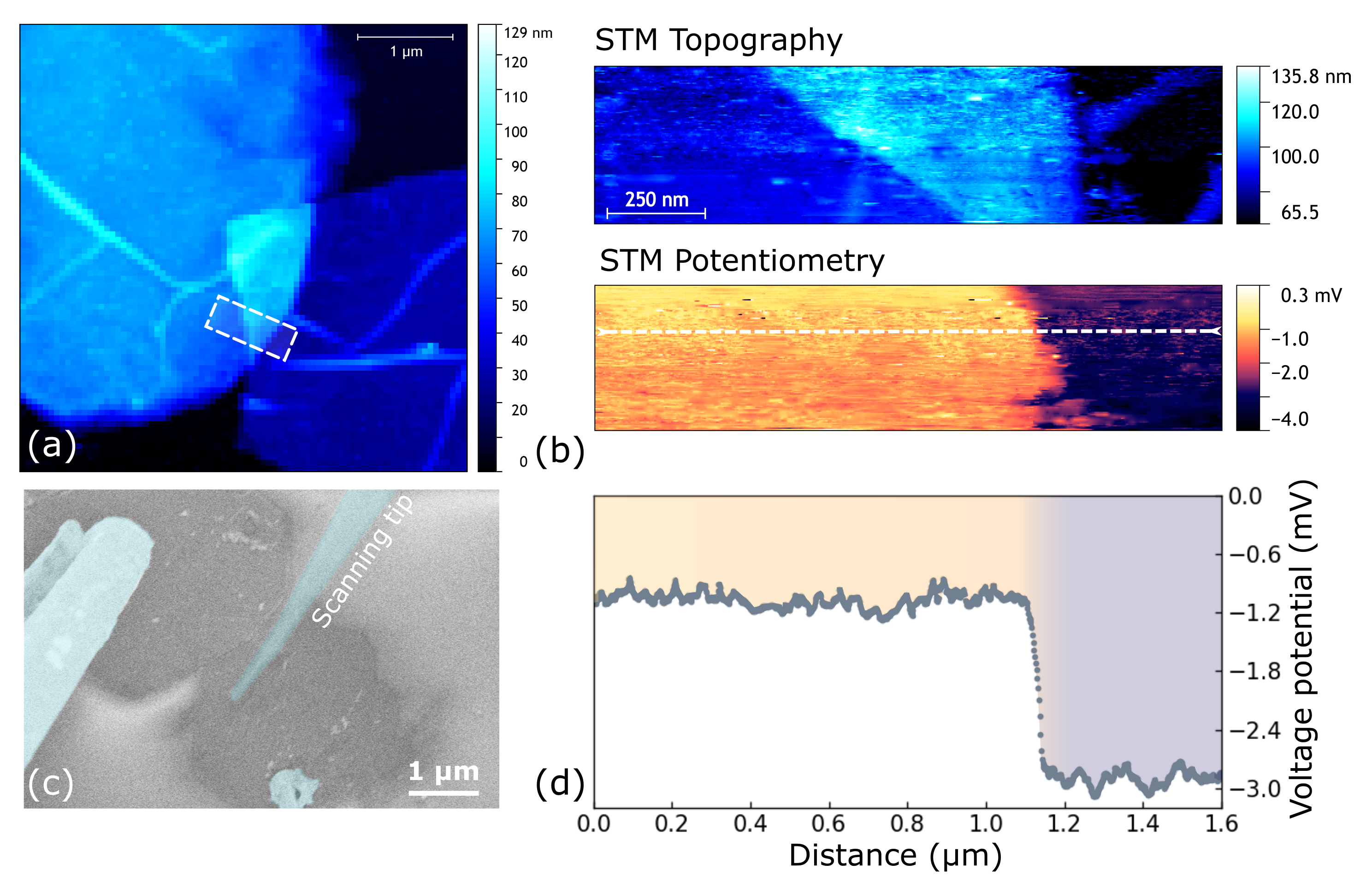}
    \caption{(a) AFM topography of the junction resistance, obtained in tapping mode. (b) STM topography and potentiometry inside the rectangle highlighted in (a). (c) SEM image of the probe configuration of the potentiometry measurement obtained with a 100 nA current between the two outer probes. (d) Profile of the potential along the white dashed line in (b).}
    \label{FIG3}
\end{figure}

As suggested by the steps shown in Figure \ref{FIG2} (f), the resistance increase appears to be sharp and very localized. To confirm this observation, Scanning Tunneling Potentiometry (STP) measurements are carried out on the same sample, at the location corresponding to the third junction, represented as $R_{J3}$ on Figure \ref{FIG2} (f). This junction was chosen as we can identify the region corresponding to a clear overlap between the flakes, with a limited amount of defects and wrinkles compared to the other junctions. STP yields a map of the local electrical potential while applying a given current on the sample with the two external STM probes \cite{Potentiometry}. This measurement configuration is shown on Figure \ref{FIG3} (a) and (c) and the obtained maps of STM topography and potential voltage are acquired simultaneously and shown on \ref{FIG3} (b). Interestingly, a sharp decrease in the measured potential is observed at the edge of the overlapping area of both flakes. This is highlighted on Figure \ref{FIG3} (d) showing a profile in the electrical potential map across the junction. On this profile, the potential is found to be approximately constant, and the voltage drop of 1.8 mV at the junction can be directly associated with the junction resistance. The resistance drop at the junction is then calculated from the voltage step, yielding a value of 18.4 k$\Omega$, which is close to the value of $R_{J3}$ measured by 4-probe measurements. The difference between both values can be explained by the higher imprecision on the voltage measurement with STP than with 4-probe measurements, but STP really gives access to local resistance variations. % The obtained value shows a good correspondence between STP and 4-probe measurements, as it is the \textcolor{blue}{same order of magnitude as $R_{J3}$.} 
Those results are in agreement with Gabbett's description of a junction-limited network, and in line with what is suggested by Kelvin probe force microscopy (KPFM) results on a \ch{MoS_2} device in operando, where it is also possible to observe sharp changes in the surface potential \cite{Gabbett}\cite{pevsic2025imaging}. The additional insight given by STP here is the ability to consider only one junction at once, with high precision of both current injection position and electrical potential measurement position.\\ %This would not be possible with KPFM without fabricating a device with contacts on the flakes of interest.

\begin{figure}[H]
    \centering
    \includegraphics[width=0.8\linewidth]{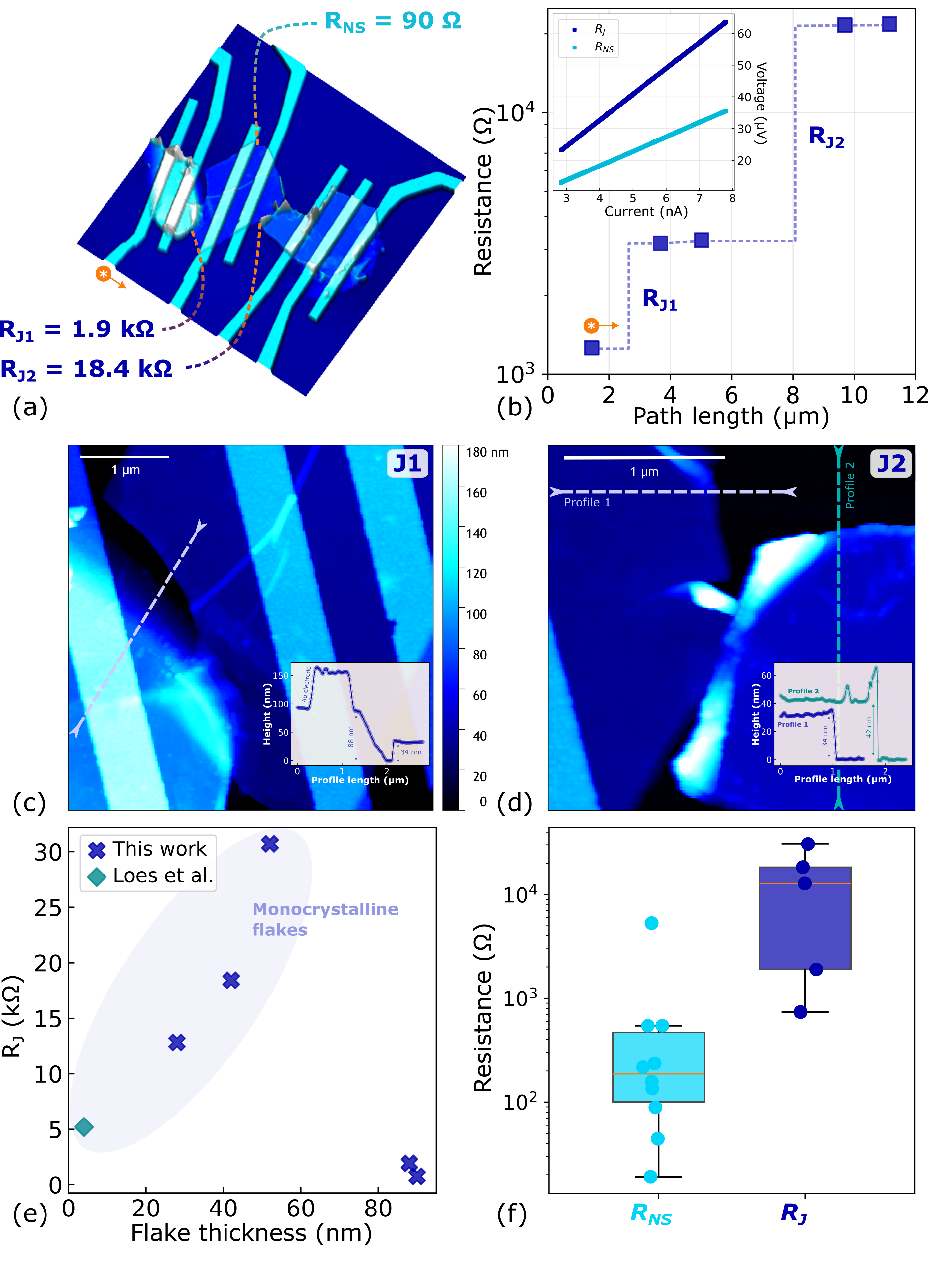}
    \caption{(a) 3D view of the AFM topography of a device with metallic contacts on \ch{Ti_3C_2T_x} flakes, obtained in tapping mode. (b) Electrical resistance along the conducting path as a function of the distance between the measuring contacts. The inset shows I-V curves acquired on one junction and on the individual flake. (c-d) Topographical details of the junctions between the flakes of the device shown in (a). The insets show height profiles extracted at the locations indicated by the dashed lines on the AFM scans. (e) Junction resistance values, along with flake thickness, for this work and the value reported in Loes et al. \cite{loes2024layer}. (f) Boxplot representing all the values obtained from fabricated devices and from 4-probe STM measurement for flake resistance ($R_{NS})$ and junction resistance ($R_J$).}
    \label{FIG4}
\end{figure}

\paragraph{Conductive path devices} %% faire ref à la figure c
To get one step closer to contacted flake networks, the next step consists in studying transport in conductive MXene paths on fabricated devices where metallic contacts are deposited directly on the flakes through a lift-off process. The contact geometry is designed to discriminate the measurements on individual flakes from those including a junction between two flakes. The AFM topography of such a device is presented on Figure \ref{FIG4} (a), which shows three individual flakes forming a conductive path (different from the one presented in Figure \ref{FIG2}), covered with 7 metallic contacts. This specific geometry allows to perform 4-contact measurements on both junctions and on individual flakes. The sample was measured in ambient conditions and exhibited increased resistance values on the segments of the conducting path where a junction is present, as shown by the graph on Figure \ref{FIG4} (b), where R$_{J1}$ = 1.9 k$\Omega$ and R$_{J2}$ = 30.7 k$\Omega$, representing the two junction resistances as indicated on Figure \ref{FIG4} (a-b). The inset of Figure \ref{FIG4} (b) shows that linear IV curves are obtained both on sections including junctions and individual flakes. The resistance ranges measured as well as the IV curves linearity are consistent with the 4-probe STM results presented above, despite a difference in the measurement conditions as the devices are measured at ambient conditions here, compared to UHV above. In both cases, it appears that junctions are the principal resistive component of the conductive path. \\
 %Also, linear IV curves are obtained both on sections including junctions and individual flakes, as presented in the inset of Figure \ref{FIG4} (b). The resistance ranges measured as well as the IV curve linearity are consistent with the 4-probe STM results presented above, despite a difference in the measurement conditions as the devices are measured at ambient conditions here, compared to UHV above. Here again, it appears that junctions are the principal resistive component of the conductive path.\\
 
\noindent Previous works have shown theoretically and experimentally that a strong resistivity anisotropy is present in \ch{Ti_3C_2T_x}, in other words the out-of-plane resistivity is about three orders of magnitude higher than the in-plane resistivity \cite{Hu_anisotropyTi3C2Tx}\cite{de2025anisotropic}. Therefore, the significant difference between $R_J$ and $R_{NS}$ can be attributed mainly to the appearance of an out-of-plane component in the transport at the junction location when the flakes are overlapping, as illustrated in the supplementary materials S3 where the junction resistance is decoupled between an out-of-plane component and the contact resistance between the flakes. This out-of-plane component becomes then the main contribution to the total resistance of the conductive path considered, and is morphology- and thickness-dependent. 
Here, a closer look at the junction morphology is provided by the AFM topography maps shown on Figure \ref{FIG4} (c-d), along with height profiles along the dashed lines represented on the AFM maps. Such precise topographical details allow to evaluate the relation between junction resistance values and junction morphology, characterized typically by flake thickness and by the overlap area between the flakes. Here, representing junction resistance values along with flake thickness (i.e. the thickness of the flake that induces the out-of-plane component of charge transport) on Figure \ref{FIG4} (e) suggests that junction resistance increases with the flake thickness, for monocrystalline flakes. Note that all of the junction resistances shown on this graph have been extracted using 4-contact measurements, either on fabricated devices or with 4-probe STM, as detailed in the supplementary information S4. An additional value from the work Loes et al. is shown on the same graph \cite{loes2024layer}. Two measurements also show a lower junction resistance for thick (> 80 nm) and disordered flakes, which have to be separated from the monocrystalline case. 
The increase in junction resistance with flake thickness observed here is consistent with the model of Gabbett and coworkers, in which the resistivity of a nanosheet network is expressed with a linear dependence on flake thickness \cite{Gabbett}. Recent experimental results on spray-cast graphene nanosheet networks also confirm this linear dependence and fit this model \cite{coleman2025decoupling}. In addition to considering flake thicknesses, AFM maps also allow to extract the overlap area for each junction. Here, junction resistance values are relatively independent of overlap area, as shown in the supplementary information S4. Defects, crystallographic orientation, intercalated species, and flake thickness have then a higher impact on junction resistance, and yield scattered values in the measurements. This is clearly seen on Figure \ref{FIG4} (f) which gathers resistance values of individual flakes ($R_{NS}$) and junctions ($R_{J}$) obtained in this work, and while the dominance of $R_J$ is clear, significant dispersion in those values is observed. %which can be attributed to morphology differences, i. e. overlap area between the flakes, flake sizes and thicknesses. 
Such an observation can be linked directly to the high variability of the overall resistivity in MXene films \cite{roadblocks}. For example, several works show that the conductivity and performances of a \ch{Ti_3C_2T_x} can be tuned as a function of the network morphology \cite{Niksan}. Guo and coworkers show that they could overcome low intrinsic conductivity in transparent conductive electrodes by increasing flake size above 10 $\mu$m, decreasing the amount of junctions in the network, therefore limiting the contribution of out-of-plane resistivity \cite{guo2023rational}. % AJouter un truc avec soft delamination et la conductivité.
Ultimately, gaining control on network morphology is likely to be a key-element to achieve a better reproducibility in MXenes network resistivity. \\

%VARIABILITÉ DANS LES ENCRES DE MXene --> RJUNCT !!! 
%   Citer papiers conductivity ? 
%DONC CCL INTERCALATION & FLAKE SIZE SUPER IMPORTANT

\paragraph{C-AFM}

\begin{figure}[H]
        \centering
        \includegraphics[width=\linewidth]{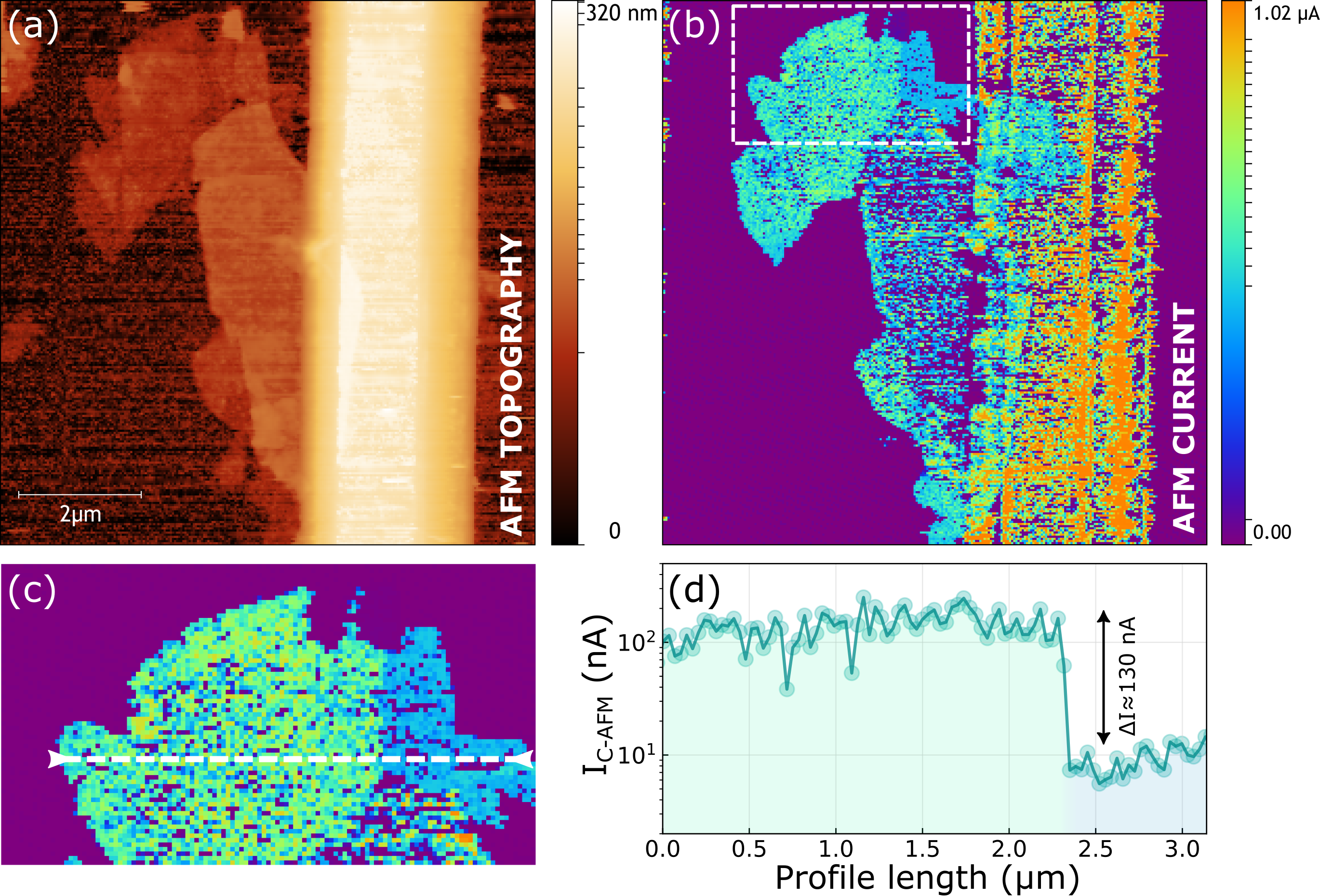}
        \caption{(a) 3D view of the AFM topography and (b) current map of \ch{Ti_3C_2T_x} flakes overlapping a metallic electrode and an insulating substrate, obtained with Conductive AFM with a 3.5 mV bias between tip and sample.  (c) Detail in the current map shown in (b). (d) C-AFM current profile along the white dotted line in (c), taken with a 5-pixel thickness (34 nm).}
        \label{FIG5}
    \end{figure}

In complement to the STP experiments, Conductive Atomic Force Microscopy (C-AFM) is performed on \ch{Ti_3C_2T_x} flakes deposited overlapping an insulating substrate and a metallic electrode, as presented on Figure \ref{FIG5} (a). This technique has similarities with STP, but this time it is performed under ambient conditions and has less stringent scan size limitations than STM, which allows to image several entire flakes at once. C-AFM allows simultaneous acquisition of a map of the sample topography and a map of the current flowing between the tip and sample, therefore allowing to image the sample conductivity changes. With this in mind, Figures \ref{FIG5} (b) and (c) show a C-AFM scan in which a junction between two flakes can be observed in a region where the C-AFM signal is stable. It is challenging to obtain quantitative findings based on the absolute current values measured in C-AFM as the 2-contact measurement (tip-sample) continuously evolves over time, tip state or humidity \cite{yuan2024effect}\cite{vazirisereshk2021time}. 
Nevertheless, the current map obtained here with C-AFM reveal that the current values measured over the flake surfaces are homogeneous, while a drastic change occurs across flake junction. To gain more precise insight, we can extract a current profile in the area of interest, as depicted on Figures \ref{FIG5} (c-d) and obtain the value of the current drop between the two flakes, which is about 130 nA. Additional C-AFM results and profiles show similar current drops at flake junctions and are presented in the supplementary information S5. These results are in excellent agreement with STP presented above and demonstrate that a MXene conductive path can be considered as isopotential domains separated by junctions, in both UHV and ambient conditions. \\

\noindent
\paragraph{Humidity response of \ch{Ti_3C_2T_x}}

\begin{figure}[H]
    \centering
    \includegraphics[width=\linewidth]{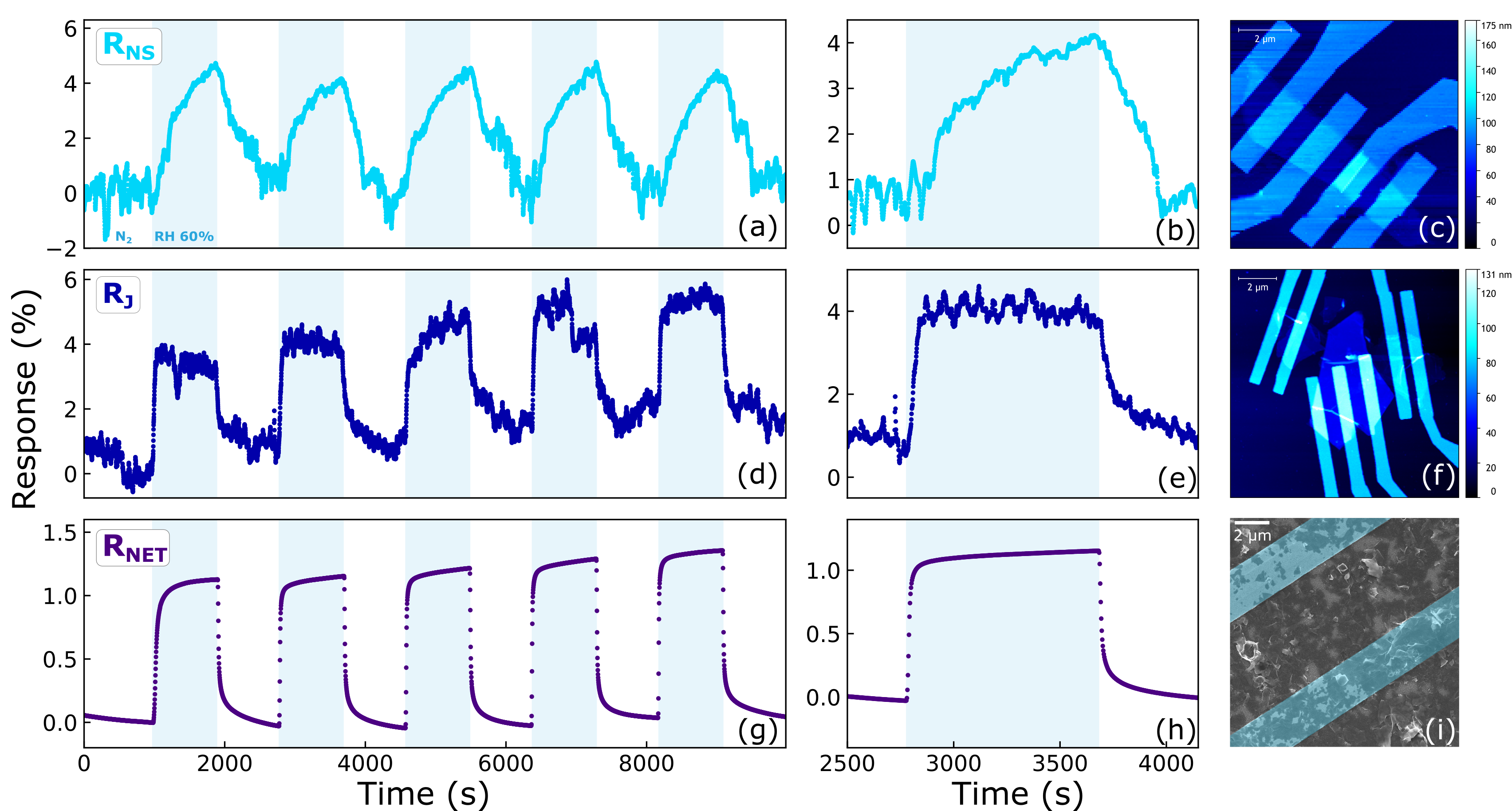}
    \caption{Response of \ch{Ti_3C_2T_x} to humidity. Response of  (a) an individual flake ($R_{NS}$ = 10.5 k$\Omega$ - 2 contacts), (d) two flakes including a junction ($R_{J}$ = 18.8 k$\Omega$ - 4 contacts), (g) a sensing layer deposited on interdigitated electrodes ($R_{NET}$ = 126 $\Omega$ - 2 contacts) to 60\% humidity in a \ch{N_2} environment. %The response is obtained as $Resp = \frac{R_{hum} - R_{N2}}{R_{N2}}$ where $R_{hum}$ is the measured resistance and $R_{N2}$ is the baseline resistance value measured after stabilization under \ch{N_2}. 
    Each humidity pulse is indicated by light blue rectangles on the graphs. (b),(e),(c) Closer view of the responses to one humidity pulse. (c), (f) AFM topography maps of the samples measured in (a) and (d) respectively. (i) SEM image of two of the digits covered with a layer of \ch{Ti_3C_2T_x}, measured in (g).}
    \label{FIG6}
\end{figure}

To illustrate the profound impact of MXene junction on MXene electronic applications, we studied the effect of humidity on the conductance of \ch{Ti_3C_2T_x}. For this, several devices involving different flake configurations have been exposed to various humidity levels in a \ch{N_2} atmosphere, in order to compare the case of an individual flake, two flakes including a junction, and a layer of networked flakes, as shown in Figures \ref{FIG6} (c,f,i). Figures \ref{FIG6} (a,d,g) display the response of each sample to several humidity pulses, after stabilization in dry \ch{N_2}. Here, the response is defined as $Resp = \frac{R_{hum} - R_{N2}}{R_{N2}}$ where $R_{hum}$ is the measured resistance and $R_{N2}$ is the baseline resistance value. A closer view of each response is shown on Figures \ref{FIG6} (b,e,h) as well as the morphology of the samples with AFM topography maps of both flake-size devices on Figures \ref{FIG6} (c,f) and a SEM image of the \ch{Ti_3C_2T_x} layer on Figure \ref{FIG6} (i). Each sample shows an increase in resistance upon humidity exposure, which corresponds to a p-type sensing behavior. This type of response is expected, as most of the reports about sensing capabilities of \ch{Ti_3C_2T_x} show a p-type response to reducing gases and humidity \cite{choi2020situ}\cite{lee2017room}\cite{kim2018metallic}\cite{wang2025ti3c2tx}. In contrast, in a recent work on layer-dependent sensing mechanism in MXenes, Loes and coworkers prove that the sensing behavior of a \ch{Ti_3C_2T_x} monolayer is intrinsically n-type \cite{loes2024layer}. They show that monolayer flakes exhibit a decrease of resistance upon exposure of reducing analytes, and that the response becomes p-type in bilayer structures due to molecular intercalation between the layers. Here, the observed all- p-type responses show that the dominant interaction with the analyte is intercalation, even for individual flakes. Therefore, a distinction has to be made between the case of the monolayer flake, the case of monocrystalline few-layer flakes, and multilayered structures that include junctions. \\

\noindent It is worth noting that the dynamics of sensing also depends on the layer number and their stacking. A faster n-type response was observed in the monolayer case by Loes et al. Here, the 12 nm-thick individual flake shows a relatively slow, p-type response. The case of the junction here can be compared with the bilayer structure in Loes, as a junction is present in both cases and shows a fast, p-type response. The visible differences in the sensing kinetics observed in Figure \ref{FIG6} can be evaluated with the response time, defined as the time to reach 90 \% of the maximal response. Here, the response time is 606 s for the individual flake, and 61 s for the junction. This one order of magnitude difference suggests that molecular intercalation occurs slower in the monocrystal than at the junction between flakes. Now, the flake-size experiments can be compared to the case of the network of flakes presented on Figure \ref{FIG6} (g). Here, fast, p-type response is observed with a response time of 44 s, which is in the same range as the case of the junction. Therefore, we conclude that the layer-scale sensing kinetics of \ch{Ti_3C_2T_x} flake networks is likely dominated by junctions.\\ %This could be investigated by further studies to evaluate how the sensing performances of MXene layers are affected by network morphology (number of junctions, flake size, flake thickness).

\noindent
In summary, we studied charge transport through MXene conductive paths by decomposing the contributions to the total electrical resistance into individual flake resistances and junction resistances. Both contributions were independently measured with 4-probe STM, C-AFM and measurements on fabricated devices with contacts patterned directly on the \ch{Ti_3C_2T_x} flakes. It was possible to establish a quantitative (more than an order of magnitude) difference between junction resistances and flake resistances, as per the model of Gabbett et al. \cite{Gabbett}, using 4-probe measurements on devices and with multiple-probe STM. Those were carried out under different atmospheres including UHV, pure \ch{N_2} at ambient pressure, dry and humid air. We could experimentally confirm that junction resistances are the main resistive component in MXene conductive paths, and that the junction resistance values are strongly morphology-dependent, as the measured junctions showed a broad dispersion in their resistance values. Additionally, we could image the junction resistances locally with STP and C-AFM, revealing a step-like change in the resistance at the junction location. With those measurements, it is possible to approximate charge transport in MXene conductive paths as through a series of isopotential flakes and a sum of junction resistances. These findings are shedding light on charge transport in MXene layers with reliable experimental validation given the multi-technique approach. Gaining such in-depth knowledge about nanoscale transport in \ch{Ti_3C_2T_x} is invaluable in the design of applications with MXene-network-based layers, typically to gain control over the conductivity of the layers deposited as well as their reactivity towards environmental changes. We illustrate the latter approach through the contrasting dynamics in the response of multilayer MXene monocrystals and of multi-junction network to humidity pulses. This can be directly applied in flexible electronics, gas sensing applications or plasmonics.

\section*{Methods}
\paragraph{\ch{Ti_3C_2T_x} synthesis}
\ch{Ti_3C_2T_x} is synthesized using the MILD (Minimally Intensive Layer Delamination) method, consisting in the etching of the 'A' layer of a MAX precursor. In this case, \ch{Ti_3AlC_2} from Carbon Ukraine is used. The MILD method relies on the use of a milder etchant than concentrated HF: \textit{in situ} HF. For this, 1.6 g of LiF salt (Sigma-Aldrich, 99\%) is stirred with 20 mL HCl 9M (Sigma-Aldrich, 37 \%) until full dissolution of the LiF. After, 1 g \ch{Ti_3AlC_2} is added slowly to the solution and kept under stirring in a Teflon beaker covered by a lid, and heated at 40 $^{\circ}$C in an oil bath for 24 hours. As the reaction between the etchant and MAX phase is strongly exothermic, it is important to add the MAX powder very progressively to avoid heatpoints that damage the crystals. The delamination step takes place after the etching and consists in washing cycles with DI water. For this, several centrifugation steps are carried out at 6000 RPM for 5 minutes and repeated until the pH level of the supernatant approaches 7. The supernatant turns black, indicating the start of the delamination, and can be collected over 2-3 additional washing cycles. \ch{Ti_3C_2T_x} is obtained by collecting the slurry above the \ch{Ti_3C_2T_x}/\ch{Ti_3AlC_2} sediment and filtering it by vacuum filtration using a 0.22 µm membrane filter. The obtained products are left to dry overnight at room temperature, separated from the membrane filter and kept in an inert atmosphere to prevent oxidation.

%TO PUT IN SUPPINFO -- SEM images

\paragraph{Samples fabrication}
%TO PUT IN SUPPINFO -- MICROSCOPE IMAGES SAMPLES BEFORE DEVICE
In order to deposit \ch{Ti_3C_2T_x} flakes on a Si/\ch{SiO_2} substrate, the dried products from the synthesis are dispersed in DI water by sonication for 1 minute. It is advised to keep the sonication time short as it can break the flakes or add defects. Deposition is performed by drop casting. The flake concentration on the substrate can be adjusted as a function of the time for which the droplet is in contact with the sample surface before being removed with a cleanroom wipe which can not touch the sample surface. Different flake concentration on the substrate are shown in the supporting information S6. The deposited flakes are then observed with optical microscopy to select the areas of interest. The samples intended for 4-probe STM are then stored under inert atmosphere, and the flakes of interest are measured with AFM in tapping mode to obtain more details about their topography.\\

\noindent
The samples intended for full device fabrication go through more process steps. The contact patterns are created with electron beam lithography, on a spin-coated 200 nm PMMA layer. Then follows a development in IPA:MIBK after which metallic layers are deposited by thermal evaporation. A 5 nm Ti adhesion layer and a 55 nm Au layer were deposited. Lift-off was carried out with acetone, and the as-fabricated devices were imaged with optical microscopy and AFM in tapping mode to validate the contact geometry. After validation, the samples were glued and wire bonded on DIL16 sockets.

\paragraph{4-probe STM}
4-probe measurements on \ch{Ti_3C_2T_x} are carried out in a multiple probe STM system, in a UHV environment, with prior preparation of the tungsten tips and the sample. The tips were placed under SEM visualization to ensure the landing of the tip on \ch{Ti_3C_2T_x} and not on the \ch{SiO_2} surface. When all four tips are in contact, it is then possible to perform four-contact measurements by applying a given current to the two external probes and measuring the voltage drop between the two inner probes. SEM visualization also allows to measure the effective distance between the probes.\\

\noindent
The multiple probe STM is also used to perform STP, in which three tips are used. Again, the two external probes apply a given current through the sample, and the third tip is scanning the surface of the sample. The scan produces a map of the sample topography and a map of the voltage potential, simultaneously acquired.

\paragraph{Measurements on fabricated devices}
All of the measurements carried out on the fabricated devices were done using one or several lock-in amplifiers (MFLI, Zurich Instruments). This configuration allowed measurement of very conductive samples while keeping low current values as the samples are quite fragile given their size. The lock-in amplifiers were used at 117 Hz with a current below 20 nA going through the samples. Those measurements are carried out in both ambient air and controlled atmosphere. For the latter, a specific setup is used, in which a gas chamber is fixed on the measuring PCB and connected to gas lines. In this case, measurements were done with nitrogen and various humidity levels, generated by a bubbler and measured by a Winsen reference sensor. The gas flux is kept constant at 1000 ml/min, using mass flow controls on the gas lines.

%%%% ALLER LIRE https://www.nature.com/articles/s41928-021-00684-9?fromPaywallRec=false %%%%
\bibliographystyle{unsrt}
\bibliography{biblio}

\section*{Data availability}
The data that support the findings of the current study are available from the corresponding author on reasonable request.

\section*{Acknowledgments}
\noindent
%B.H. (senior research associate) acknowledges support from the F.R.S.-FNRS. The authors are grateful to the Walloon region for funding this research through SIAHNA and ASPCIN projects. The authors also acknowledge financial support by the ARC project DREAMS (21/26.116).
B.H. (senior research associate) is supported by the F.R.S.-FNRS. The authors are grateful to the Walloon region for funding this research through SIAHNA project. The authors also acknowledge
financial support from the ARC project DREAMS (21/26.116). The authors also thank the IEMN-PCMP-PCP platform and Renatech network. The authors thank Cécile D'Haese and Benoît Hubert for their support, as well as VOCSens team for the collaboration and technical support.

%\section{Author contributions}

\section*{Competing interests}
The authors declare no competing interests.
\end{document}